\documentclass[11pt]{article}

\usepackage[a4paper, total={6.5in, 8.5in}]{geometry}

\usepackage[utf8]{inputenc} 
\usepackage{subcaption}
\usepackage{xcolor}
\usepackage[T1]{fontenc}    
\usepackage{hyperref}       
\usepackage{url}            
\usepackage{booktabs}       
\usepackage{amsfonts}       
\usepackage{nicefrac}       
\usepackage{microtype}      
\usepackage{lipsum}
\usepackage{fancyhdr}       
\usepackage{graphicx}       
\usepackage{amsmath}
\usepackage{physics}
\usepackage{float}

\graphicspath{{media/}}     

\title{Evaluation of derivatives using approximate generalized parameter shift rule

}

\author{
  Vytautas Abramavicius,
  Evan Philip,
  Kaonan Micadei,
  Charles Moussa,\\
  Mario Dagrada,
  Vincent E. Elfving,
  Panagiotis Barkoutsos,
  Roland Guichard
  \\
  Pasqal, 24 Av. Emile Baudot, 91120 Palaiseau, France
}
\date{}

\begin{document}
\maketitle

\begin{abstract}
Parameter shift rules are instrumental for derivatives estimation in a wide range of quantum algorithms, especially in the context of Quantum Machine Learning. Application of single-gap parameter shift rule is often not possible in algorithms running on noisy intermediate-scale quantum (NISQ) hardware due to noise effects and interaction between device qubits. In such cases, generalized parameter shift rules must be applied yet are computationally expensive for larger systems. In this paper we present the approximate generalized parameter rule (aGPSR) that can handle arbitrary device Hamiltonians and provides an accurate derivative estimation while significantly reducing the computational requirements. When applying aGPSR for a variational quantum eigensolver test case ranging from $3$ to $6$ qubits, the number of expectation calls is reduced by a factor ranging from $7$ to $504$ while reaching the exact same target energy, demonstrating its huge computational savings capabilities.
\end{abstract}


\section{Introduction}
Derivative calculations for quantum circuit outputs with respect to circuit parameters is essential to implement a breadth of quantum algorithms, especially in the field of Quantum Machine Learning (QML). Notable examples include algorithms like Differentiable Quantum Circuit (DQC) \cite{kyriienko2021solving} and Quantum Extremal Learning (QEL) \cite{varsamopoulos2022quantum}. Analytical or automatic differentiation (AD) methods are usually favored over numerical differentiation methods, \textit{e.g.} based on finite differences, even in classical machine learning due to their robustness against errors. Quantum computation suffer fundamentally from shot noise, therefore numerical differentiation is error prone and unfeasible on noisy intermediate-scale quantum (NISQ) hardware with limited shot budget. On quantum hardware where qubit interactions can be limited or turned off (superconducting or ion-trapped), control parameters are generally rotation angles in single-qubit gates. Under these settings, the parameter shift rule (PSR) \cite{PhysRevA.98.032309, PhysRevA.99.032331} has become the \textit{de facto} method for calculating derivatives. It has also been experimentally demonstrated to be more robust against noise \cite{ishiyama2022noise}. 

Usual PSR is valid where parameters control gates that can be described by involutary and idempotent generators having spectrum consisting of unique values $\pm \lambda$. However, not all parametrized quantum circuits can be represented by such generators. Notably, on neutral atom platforms such as Pasqal's, any non-trivial configuration for a parameterized quantum circuit exhibits a background Rydberg interaction between atoms that cannot be turned off. In such cases, generators governing the dynamics of the quantum circuit are neither involutary nor idempotent, thus one has to rely on the generalized parameter shift rule (GPSR) \cite{gpsr2021} to calculate derivatives. 

The derivative expression in \cite{gpsr2021} is free from any approximations, hence mathematically exact. It provides a method to calculate the exact derivative with respect to parameters represented by an arbitrary generator with a rich spectrum of eigenvalues. Moreover, it remains robust against noise by allowing tailored parameter shift values $\delta_s$ to minimize the variance of derivative estimation. Since the generator of a $N$-qubit quantum circuit has $2^N$ eigenvalues and (at most) $S=\frac{2^N (2^N-1)}{2}$ unique spectral gaps, GPSR requires making measurements at $2S$ different values $x\pm\delta_s$ of the parameter to estimate the derivative at any point $x$. For instance, derivative calculations of a $N=5$ qubit circuit on neutral atom hardware cannot be performed with ordinary PSR, and its generalized version has to applied. However, even with a 1000-shot budget, merely a single estimate at each of the $2S=992$ shifted points $x\pm\delta_s$ required by GPSR will be possible, rendering it practically useless for the calculation of the derivative. 

To solve this exponential scaling problem, we propose aGPSR (approximate GPSR) as a method of estimating derivative of a function spawned by an arbitrary generator having a non-trivial spectrum of eigenvalues in a limited shot budget setting. 

The structure of the paper is as follows. Section~\ref{background} outlines the theoretical foundations of aGPSR for derivative estimation. We then perform an analysis of the (minimization of) variance of the derivative estimation in Section~\ref{variance_est}. Finally, we demonstrate the usage of aGPSR and its potential in computational savings in Section~\ref{results}.

\section{GPSR and Approximate GPSR}
\label{background}

In this section, we first outline the background and notations of GPSR to introduce aGPSR. 

\subsection{Derivation with GPSR}

Let us first outline the necessary background and notations following the GPSR work\cite{gpsr2021}. Let $f(x)$ be a function of a parameter $x$ defined as the quantum expectation of a Hermitian cost operator $\hat{C}$
\begin{equation}
f\left(x\right)=\left\langle\psi_0\right|\hat{U}^{\dagger}(x)\hat{C}\hat{U}(x)\left|\psi_0\right\rangle, \label{eq:f(x)}
\end{equation}
where $\left|\psi_0\right\rangle$ is some initial state and $\hat{U}(x)={\rm exp}\left(\frac{-ix}{2}\hat{G}\right)$ is a unitary evolution operator with its dynamics determined by the Hermitian generator $\hat{G}$. The derivative of $f(x)$ is given by 
\begin{equation}
\frac{{\rm d}f\left(x\right)}{{\rm d}x}=\overset{S}{\underset{s=1}{\sum}}\Delta_{s}R_{s}, \label{dfdx_definition}
\end{equation}
where $S$ is the number of unique spectral gaps of generator $\hat{G}$ denoted $\{\Delta_s\}_{s=1}^S$. Parameters $R_s$, necessary for derivative evaluation, can be obtained by solving the following system of linear equations:
\begin{equation}
\begin{cases}
F_{1} & =4\overset{S}{\underset{s=1}{\sum}}{\rm sin}\left(\frac{\delta_{1}\Delta_{s}}{2}\right)R_{s},\\
F_{2} & =4\overset{S}{\underset{s=1}{\sum}}{\rm sin}\left(\frac{\delta_{2}\Delta_{s}}{2}\right)R_{s},\\
 & ...\\
F_{S} & =4\overset{S}{\underset{s=1}{\sum}}{\rm sin}\left(\frac{\delta_{S}\Delta_{s}}{2}\right)R_{s}.
\end{cases}\label{Rs_eq_system}
\end{equation}
Here we denoted $F_s=f(x+\delta_s)-f(x-\delta_s)$ as the difference between shifted functions $f(x)$. In a more compact way the linear equation system can be written as a matrix equation
\begin{equation}
\mathbf{F}^{S\times 1}=\mathbb{M}^{S\times S}\mathbf{R}^{S\times 1},\label{matrix_eq_system}
\end{equation}
where $\mathbb{M}^{S\times S}$ is a $S\times S$ matrix with $\mathbb{M}^{S\times S}_{ij}=4{\rm sin}\left(\frac{\delta_i\Delta_j}{2}\right)$ and $\mathbf{F}$ and $\mathbf{R}$ are vectors containing $F_s$ and $R_s$ components respectively.

\subsection{Approximate GPSR} \label{subsec: deriv}

However, as stated previously, the number of equations involved in the previous system \ref{matrix_eq_system} can be prohibitive. Approximate GPSR aims at obtaining the derivative expression when the number of equations in this system is reduced to some $K<S$ and equation similar to Eq. (\ref{dfdx_definition}) is applied to estimate the derivative. First, we redefine the unknown values $R_s$ to $R_s^{\prime}$ on the rhs of system (\ref{Rs_eq_system}) and introduce new spectral gaps (let us refer to them as \guillemotleft pseudo-gaps\guillemotright) $\{\gamma_s\}_{s=1}^K$ that in general case do not coincide with the original set of generator spectral gaps $\{\Delta_s\}_{s=1}^S$. We propose in the later subsection~\ref{subsec: gap-select} a few strategies to select them. With these notations the $s$-th equation of system \ref{Rs_eq_system} reads:
\begin{equation}
F_s=4\overset{K}{\underset{k=1}{\sum}}{\rm sin}\left(\frac{\delta_{s}\gamma_{k}}{2}\right)R_{k}^{\prime}, \label{Rs_eq_gamma}
\end{equation}
where now $\mathbb{M}^{K\times K}_{ij}=4{\rm sin}\left(\frac{\delta_i\gamma_j}{2}\right)$. The corresponding definition of the approximated derivative, thus is given by
\begin{equation}
\left.\frac{{\rm d}f\left(x\right)}{{\rm d}x}\right|_{K}=\overset{K}{\underset{s=1}{\sum}}\gamma_{s}R_{s}^{\prime}. \label{dfdx_definition_gamma}
\end{equation}
Using matrix notation the "effective" linear system now can be written as
\begin{equation}
\mathbf{F}^{K\times 1}=\mathbb{M}^{K\times K}\mathbf{R^{\prime}}^{K\times 1}.\label{matrix_eq_system_eff}
\end{equation}

The general solutions for $R_s^{\prime}$ of Eq. \ref{matrix_eq_system_eff} can be obtained using the Cramer's rule:
\begin{equation}
R^{\prime}_k=\frac{{\rm det}(\mathbb{M}^{K\times K}_{(k)})}{{\rm det}(\mathbb{M}^{K\times K})},\label{KxK_eq_cramers}
\end{equation}
where $\mathbb{M}^{K\times K}_{(k)}$ is the matrix formed by replacing $k$-th column of $\mathbb{M}^{K\times K}$ matrix with the vector $\mathbf{F}^{K\times 1}$. We can simplify this solution by using the explicit components of the vector $\mathbf{F}^{K\times 1}$ with original set of spectral gaps $\{\Delta_s\}_{s=1}^S$ given by Eq. \ref{Rs_eq_system}. Thus, for $R^{\prime}_1$ when, e. g., $K=2$ we get
\begin{equation}
R^{\prime}_1=\frac{\overset{S}{\underset{s=1}{\sum}}R_s\begin{vmatrix}{\rm sin}\left(\frac{\delta_{1}\Delta_{s}}{2}\right) & {\rm sin}\left(\frac{\delta_{1}\gamma_{2}}{2}\right)\\
{\rm sin}\left(\frac{\delta_{2}\Delta_{s}}{2}\right) & {\rm sin}\left(\frac{\delta_{2}\gamma_{2}}{2}\right)
\end{vmatrix}}{\begin{vmatrix}{\rm sin}\left(\frac{\delta_{1}\gamma_{1}}{2}\right) & {\rm sin}\left(\frac{\delta_{1}\gamma_{2}}{2}\right)\\
{\rm sin}\left(\frac{\delta_{2}\gamma_{1}}{2}\right) & {\rm sin}\left(\frac{\delta_{2}\gamma_{2}}{2}\right)
\end{vmatrix}}=\overset{S}{\underset{s=1}{\sum}}\eta_{1s}(\Delta_s, \gamma_1,\gamma_2,\delta_1,\delta_2)R_s,\label{2x2_eq_cramers_R1}
\end{equation}
where we explicitly denoted the determinant of the matrix with $|M_{ij}|$. The sum in the expression above comes from the multilinearity property of determinants stating that if a column of matrix can be written as a sum of column vectors then the determinant of such matrix is equal to the sum of determinants with the corresponding column substituted with each of the vectors. Note that each function $\eta_{1s}(\Delta_s, \gamma_1,\gamma_2,\delta_1,\delta_2)$ only depends on the corresponding "real" spectral gap $\Delta_s$ and not on other $\Delta$ values. For example, for $K=2$ and some $1\leq k \leq K$ the explicit form of function $\eta_{ks}(\Delta_s, \gamma_1,\gamma_2,\delta_1,\delta_2)$ reads
\begin{equation}
\eta_{ks}(\Delta_s, \gamma_1,\gamma_2,\delta_1,\delta_2) = \frac{{\rm sin}\left(\frac{\delta_{1}\gamma_{2}}{2}\right){\rm sin}\left(\frac{\delta_{2}\Delta_{s}}{2}\right)-{\rm sin}\left(\frac{\delta_{2}\gamma_{2}}{2}\right){\rm sin}\left(\frac{\delta_{1}\Delta_{s}}{2}\right)}{{\rm sin}\left(\frac{\delta_{1}\gamma_{2}}{2}\right){\rm sin}\left(\frac{\delta_{2}\gamma_{1}}{2}\right)-{\rm sin}\left(\frac{\delta_{1}\gamma_{1}}{2}\right){\rm sin}\left(\frac{\delta_{2}\gamma_{2}}{2}\right)}. \label{explicit_eta_K=2}
\end{equation}

The results of Eqs. \ref{KxK_eq_cramers} and \ref{2x2_eq_cramers_R1} can be straightforwardly generalized to give solutions for any system containing any number $K$ of equations:
\begin{equation}
R^{\prime}_k=\overset{S}{\underset{s=1}{\sum}}\eta_{ks}(\Delta_s, \{\gamma_i\}_{i=1}^K,\{\delta_i\}_{i=1}^K)R_s=\overset{S}{\underset{s=1}{\sum}}\frac{{\rm det}(\mathbb{A}^{K\times K}_{(ks)})}{{\rm det}(\mathbb{M}^{K\times K})}R_s. \label{KxK_eq_cramers_Rk}
\end{equation}
Here we introduced matrices $\mathbb{A}^{K\times K}_{(ks)}$ where the $k$-th column contains the vector $\mathbf{B}^{K\times 1}_{(s)}$ composed of elements $\mathbf{B}^{K\times 1}_{si}=4{\rm sin}\left(\frac{\delta_i\Delta_s}{2}\right)$ meaning that $\mathbf{F}^{K\times 1}=\sum_{s=1}^S\mathbf{B}^{K\times 1}_{(s)}R_s$. Note that functions $\eta_{ks}(\Delta_s, \{\gamma_i\}_{i=1}^K,\{\delta_i\}_{i=1}^K)$ above are given by ratios of determinants, thus their explicit forms contain combinations of sine functions similarly to Eq. \ref{explicit_eta_K=2}. Plugging expression \ref{KxK_eq_cramers_Rk} into definition given in Eq. \ref{dfdx_definition_gamma} we can  write the approximate derivative as
\begin{eqnarray}
\left.\frac{{\rm d}f\left(x\right)}{{\rm d}x}\right|_{K} & = & \overset{K}{\underset{k=1}{\sum}}\gamma_{k}R^{\prime}_{k}=\overset{K}{\underset{k=1}{\sum}}\overset{S}{\underset{s=1}{\sum}}\gamma_k\eta_{ks}(\Delta_s, \{\gamma_i\}_{i=1}^K,\{\delta_i\}_{i=1}^K)R_s \nonumber \\
& = & \overset{S}{\underset{s=1}{\sum}}\xi_s\left(\Delta_s, \{\gamma_i\}_{i=1}^K,\{\delta_i\}_{i=1}^K\right) R_{s}. \label{dfdx_est_Keq_explicit}
\end{eqnarray}
where we introduced new functions 
\begin{equation}
\xi_s\left(\Delta_s,\{\gamma_i\}_{i=1}^K,\{\delta_i\}_{i=1}^K\right)=\overset{K}{\underset{k=1}{\sum}}\gamma_k\eta_{ks}(\Delta_s, \{\gamma_i\}_{i=1}^K,\{\delta_i\}_{i=1}^K).\label{xi_func_def}
\end{equation}

To examine the behavior of the functions $\xi_s\left(\Delta_s,\{\gamma_i\}_{i=1}^K,\{\delta_i\}_{i=1}^K\right)$, let us first rewrite angle shifts as products $\delta_s\equiv\alpha\delta^{\prime}_s$ with $\alpha$ being a small parameter that leads to $\xi_s\left(\Delta_s,\{\gamma_i\}_{i=1}^K,\{\delta_i\}_{i=1}^K\right)\Rightarrow\xi_s\left(\alpha, \Delta_s,\{\gamma_i\}_{i=1}^K,\{\delta_i^{\prime}\}_{i=1}^K\right)$. 
We can also notice that functions $\xi_s$ enter Eq. \ref{dfdx_est_Keq_explicit} in similar way as spectral gaps $\Delta_s$. Since $\xi_s$ is a function of parameter $\alpha$ we can expand the function with respect to it. In practice for $K=2$ using Eq. \ref{explicit_eta_K=2} for functions $\eta_{ks}$ together with Eq. \ref{xi_func_def} we obtain the following expansion:
\begin{equation}
\xi_s = \Delta_s+\alpha^4\frac{\delta_1^2\delta_2^2\Delta_s(\gamma_1^2-\Delta_s^2)(\Delta_s^2-\gamma_2^2)}{1920}+\mathcal{O}\left(\alpha^{6}\right).\label{xi_expansion_K=2}
\end{equation}
This expression shows that $\xi_s$ up to the $4$-th order in $\alpha$ is equal to the $s$-th spectral gap. Thus, if $\alpha$ is sufficiently small and, consequently, the angle shifts $\delta_1$ and $\delta_2$ are small as well, it is sufficient to use only $K=2$ equations to estimate the derivative while introducing only $4$-th order errors. However, as $\alpha$ gets bigger $\mathcal{O}(\alpha^{4})$ and higher order terms might get significant and we might need a third or even more equations to estimate the derivative more accurately. Repeating the same expansion procedure for $K=1$ and $K=3$ we get respectively
\begin{eqnarray}
\xi_{s} & = & \Delta_{s}+\mathcal{O}\left(\alpha^{2}\right),\;K=1\nonumber\\
\xi_{s} & = & \Delta_{s}+\mathcal{O}\left(\alpha^{6}\right),\;K=3\label{xi_expansion_K=1,3}
\end{eqnarray}
We can notice that with more equations included in the linear system, the difference $\Delta_s-\xi_s$ becomes smaller, more specifically $\Delta_s-\xi_s=\mathcal{O}\left(\alpha^{2K}\right)$. Thus, Eq. \ref{dfdx_definition_gamma} can be written as
\begin{equation}
\left.\frac{{\rm d}f\left(x\right)}{{\rm d}x}\right|_{K}=\overset{S}{\underset{s=1}{\sum}}\left(\Delta_{s}+\mathcal{O}\left(\alpha^{2K}\right)\right)R_{s}.\label{dfdx_est_Keq_err}
\end{equation}
Note that this equation contains the dependence on the set of arbitrary values $\{\gamma_s\}_{s=1}^K$ only in the higher order error terms of approximation of spectral gaps $\Delta_s$. If we disregard these error terms Eq. \ref{dfdx_est_Keq_err} actually coincides with the exact expression of derivative in full GPSR form \ref{dfdx_definition}. This means that we can in principle use any values $\{\gamma_s\}_{s=1}^K$ to solve the linear equation system \ref{Rs_eq_gamma} and still obtain an accurate approximation of the real derivative as given in Eq. \ref{dfdx_definition}.

We can also notice that terms involving $\alpha$ appear in high-order terms for large values of $K$ thus, these error terms are sufficiently small for larger values of $\alpha$ and for larger angle shifts $\delta_s$ as well. Consequently, the more equations are included in the linear system, the larger angle shift can be used to retain similar accuracy of derivative estimation. To prove this more rigorously we can use mathematical induction principle. Since we have shown that $\xi_s\approx\Delta_s$ for low number of equations, let us assume that such relation holds for $K=S-1$, thus the error is $\mathcal{O}\left(\alpha^{2S-2}\right)$. Thus, for $K=S$ the error should be on the order or smaller than $\mathcal{O}\left(\alpha^{2S}\right)$. Since in this case we solve the full equation system we know that the prefactor of each $R_s$ is equal to $\Delta_s$ exactly or equivalently the error is infinitely small - $\mathcal{O}\left(\alpha^{N}\right)$ with $N\rightarrow\infty$. Hence, the expansion is valid for $K=S$, therefore for $K=S-1$ as well. This way the assumed relation of error magnitude holds for any $K$.

Having outlined general formulations for aGPSR, we follow up by analysing the relationship between error terms in Eq. \ref{dfdx_est_Keq_err} and 

\subsection{Error terms analysis and gap selection}  \label{subsec: gap-select}

Let us investigate the error terms in Eq. \ref{dfdx_est_Keq_err} in more detail. To do that, we can write out for $K=1,2,3$ the explicit error terms for up to $\alpha=6$ order, thus giving 3 error terms for each $K$. For $K=1$ we have
\begin{eqnarray}
\left.\frac{{\rm d}f\left(x\right)}{{\rm d}x}\right|_{1} & = & \overset{S}{\underset{s=1}{\sum}}R_s\left(\Delta_{s} + \alpha^2 \frac{\delta_1^2\Delta_s(\gamma_1^2-\Delta_s^2)}{24}\right. \nonumber\\
 & + & \alpha^4\frac{\delta_1^4\Delta_s(\gamma_1^2-\Delta_s^2)(7\gamma_1^2-3\Delta_s^2)}{5760} \label{dfdx_K=1_gamma_err}\\
& + & \left.\alpha^6\frac{\delta_1^6\Delta_s(\gamma_1^2-\Delta_s^2)(31\gamma_1^4-18\gamma_1^2\Delta_s^2+3\Delta_s^4)}{967680}+\mathcal{O}\left(\alpha^{8}\right)\right), \nonumber
\end{eqnarray}
for $K=2$ we have
\begin{eqnarray}
\left.\frac{{\rm d}f\left(x\right)}{{\rm d}x}\right|_{2} & = & \overset{S}{\underset{s=1}{\sum}}R_s\left(\Delta_{s} + \alpha^4\frac{\delta_1^2\delta_2^2\Delta_s(\gamma_1^2-\Delta_s^2)(\Delta_s^2-\gamma_2^2)}{1920}\right. \nonumber\\
& - & \alpha^6\frac{\delta_1^2\delta_2^2(\delta_1^2+\delta_2^2)\Delta_s(\gamma_1^2-\Delta_s^2)(\Delta_s^2-\gamma_2^2)(10\Delta_s^2-11(\gamma_1^2+\gamma_2^2))}{3225600} \nonumber\\
& + & \left. \mathcal{O}\left(\alpha^{8}\right)\right), \label{dfdx_K=2_gamma_err}
\end{eqnarray}
and finally, for $K=3$ we obtain
\begin{eqnarray}
\left.\frac{{\rm d}f\left(x\right)}{{\rm d}x}\right|_{3} & = & \overset{S}{\underset{s=1}{\sum}}R_s\left(\Delta_{s} + \alpha^6\frac{\delta_1^2\delta_2^2\delta_3^2\Delta_s(\gamma_1^2-\Delta_s^2)(\gamma_2^2-\Delta_s^2)(\gamma_3^2-\Delta_s^2)}{322560}\right. \nonumber\\
& + & \left. \mathcal{O}\left(\alpha^{8}\right)\right). \label{dfdx_K=3_gamma_err}
\end{eqnarray}

From the expressions above we can notice that each error term contains products of form $(\gamma_k^2-\Delta_s^2)$ where $k\in\{1,2,3\}$. This means that if we choose some $\gamma_k=\Delta_s$, the corresponding error term vanishes. This observation naturally leads to conclusion that when we select to include $K=S$ equations in the linear system \ref{Rs_eq_system} and choose all of $\gamma_k$ to coincide with the actual set of spectral gaps $\{\Delta_s\}_{s=1}^S$ we recover original exact derivative as given by GPSR since all the error terms vanish. 

The form of error terms in Eqs. \ref{dfdx_K=1_gamma_err} - \ref{dfdx_K=3_gamma_err} also suggests that even if we approximate the derivative using only $K<S$ equations in the linear system \ref{Rs_eq_gamma}, we can still find some set of values $\{\gamma_k\}_{k=1}^K$ that minimize the error terms. Intuitively, this happens when the $\gamma_k$ values are close to as many actual spectral gaps as possible, thus reducing the magnitude of the error terms and pushing the sum in parentheses multiplied by the corresponding $R_s$ closer to $\Delta_s$. Let us investigate the error terms in Eqs. \ref{dfdx_K=1_gamma_err} - \ref{dfdx_K=3_gamma_err} in greater detail.

First, we denote for some $K$ the error term of order $m=2K$ as $\varepsilon_{K,m}\left(\Delta_s, \{\gamma_k\}_{k=1}^K, \{\delta_k\}_{k=1}^K\right)$. Thus, when $K$ equations are used, recall from subsection \ref{subsec: deriv} that in the aGPSR algorithm each spectral gap $\Delta_s$ is approximated as 
\begin{equation}
\Delta_s \approx \xi_s\left(\Delta_s,\{\gamma_k\}_{k=1}^K,\{\delta_k\}_{k=1}^K\right) = \Delta_s + \sum_{i=K}^{\infty}\varepsilon_{K,m}\left(\Delta_s,\{\gamma_k\}_{k=1}^K,\{\delta_k\}_{k=1}^K\right). \label{gap_approximation}
\end{equation}
We know that the sum in equation above vanishes when $K=S$: we use all the spectral gaps of the generator. However, it is important to deduce the closed form of error term series because it allows to quantify the total error of implicit estimation of spectral gaps that results from solving the restricted form of the linear system \ref{Rs_eq_system}. This in turn presents a way to calculate which set of gaps results in the smallest error of $\{\Delta_s\}_{s=1}^S$ estimation, thus finding the optimal set of values $\{\gamma_k\}_{k=1}^K$ that also minimizes the estimation error of the derivative itself. 

The error term series can be calculated in the following way. From Eq. \ref{KxK_eq_cramers_Rk} we see that function $\eta_{ks}(\Delta_s, \{\gamma_i\}_{i=1}^K,\{\delta_i\}_{i=1}^K)$ is given by
\begin{equation}
\eta_{ks}(\Delta_s, \{\gamma_i\}_{i=1}^K,\{\delta_i\}_{i=1}^K) = \frac{{\rm det}(\mathbb{A}^{K\times K}_{(ks)})}{{\rm det}(\mathbb{M}^{K\times K})}, \label{eta_s_func_def_determ}
\end{equation}
thus the function $\xi_s$ approximating the spectral gap $\Delta_s$ according to Eq. \ref{xi_func_def} is expressed as
\begin{equation}
\xi_s\left(\Delta_s,\{\gamma_k\}_{k=1}^K,\{\delta_k\}_{k=1}^K\right) = \overset{K}{\underset{k=1}{\sum}}\gamma_k\frac{{\rm det}(\mathbb{A}^{K\times K}_{(ks)})}{{\rm det}(\mathbb{M}^{K\times K})}. \label{xi_func_def_determinant}
\end{equation}
These functions $\xi_s$ are precisely the closed-form expressions of the series contained in the parenthesis of Eqs. \ref{dfdx_K=1_gamma_err} - \ref{dfdx_K=3_gamma_err}. Thus, for some number $K$ of equations contained in the linear system \ref{Rs_eq_gamma} we can introduce the full error function $Q_{K}$ measuring the deviation of the estimated spectral gap $\xi_s$ from the true spectral gap $\Delta_s$:
\begin{equation}
Q_K\left(\Delta_s,\{\gamma_k\}_{k=1}^K,\{\delta_k\}_{k=1}^K\right) = \xi_s\left(\Delta_s,\{\gamma_k\}_{k=1}^K,\{\delta_k\}_{k=1}^K\right) - \Delta_s. \label{err_func_def}
\end{equation}

\begin{figure}[htbp]
    \centering
    \includegraphics[width=0.8\textwidth]{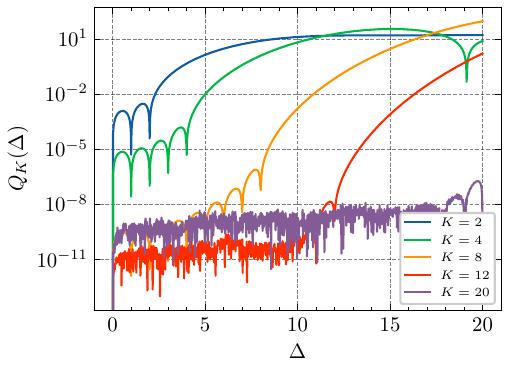}
    \caption{Error functions $Q_K(\Delta)$ for different values of $K$.}
    \label{fig:err-func}
\end{figure}

We see that the full error function $Q_K$ in principle can be calculated analytically by estimating ratios of determinants of $K\times K$ matrices. However, for large $K$ values analytical estimation of determinants requires a lot computational resources and quickly becomes unfeasible. This problem can be avoided by recalling that in Eq. \ref{xi_func_def_determinant} ratios of determinants are defined as functions $\eta_{ks}$, according to Eq. \ref{KxK_eq_cramers_Rk}, that in turn can be interpreted as solutions of the following linear system of equations:
\begin{equation}
\begin{cases}
{\rm sin}\left(\frac{\delta_{1}\Delta_{s}}{2}\right) & =\overset{K}{\underset{k=1}{\sum}}{\rm sin}\left(\frac{\delta_{1}\gamma_{k}}{2}\right)\eta_{ks},\\
{\rm sin}\left(\frac{\delta_{2}\Delta_{s}}{2}\right) & =\overset{K}{\underset{k=1}{\sum}}{\rm sin}\left(\frac{\delta_{2}\gamma_{k}}{2}\right)\eta_{ks},\\
 & ...\\
{\rm sin}\left(\frac{\delta_{K}\Delta_{s}}{2}\right) & =\overset{K}{\underset{k=1}{\sum}}{\rm sin}\left(\frac{\delta_{K}\gamma_{k}}{2}\right)\eta_{ks}.
\end{cases}\label{eta_eq_system}
\end{equation}
In this way, we can solve the system \ref{eta_eq_system} for some set of parameter values. \\ $\{\Delta_s,\{\gamma_k\}_{k=1}^K,\{\delta_k\}_{k=1}^K\}$ and obtain the set of solutions $\{\eta_{ks}\}_{k=1}^K$ that then allow to write $Q_K$ as
\begin{equation}
Q_K\left(\Delta,\{\gamma_k\}_{k=1}^K,\{\delta_k\}_{k=1}^K\right) = \overset{K}{\underset{k=1}{\sum}}\gamma_k\eta_{k}\left(\Delta,\{\gamma_k\}_{k=1}^K,\{\delta_k\}_{k=1}^K\right) - \Delta. \label{err_func_final}
\end{equation}
In this expression for $Q_K$, we dropped the dependence on index $s$ since the form of the error function is identical for every spectral gap $\Delta_s$. An example of $Q_K$ functions calculated for different numbers $K$ is presented in Fig. \ref{fig:err-func}. 

Here, the pseudo-gaps are chosen as $\gamma_k \in \{\epsilon, 1, 2, ... K-1\}$ where $\epsilon=0.1$ and shifts $\delta_k$ are equidistantly sampled from the interval $[\frac{\pi}{4}, \frac{\pi}{2}]$. The behavior of the function $Q_K$ in Fig. \ref{fig:err-func} shows that increasing the number $K$ leads to a wider interval of spectral gap values for which $Q_K$ is orders of magnitude smaller than elsewhere. This means that if for some generator $\hat{G}$ the spectral gaps are mainly distributed in this low-error region, the resulting derivative estimation will be more accurate. 

\begin{figure}[htbp]
    \centering
    \includegraphics[width=0.8\textwidth]{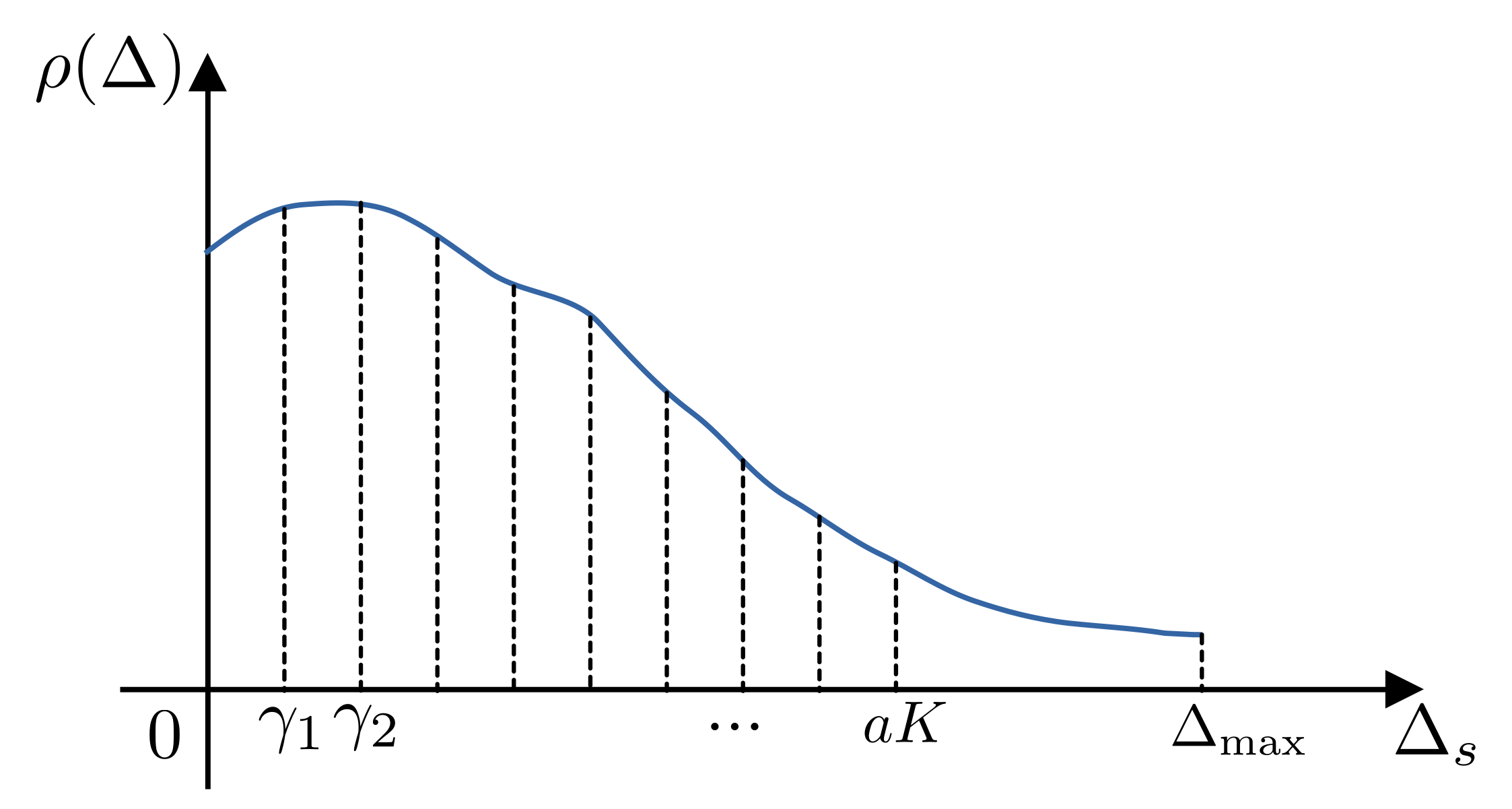}
    \caption{Illustration of uniform sampling procedure of pseudo-gaps $\gamma_k$ in the interval $[0, \Delta_{\rm{max}}]$ with generator's $\hat{G}$ gap distribution given by $\rho(\Delta)$.}
    \label{fig:gap-selection}
\end{figure}

Let us explore the interplay between the spectrum $\rho(\Delta)$ of the generator $\hat{G}$ and the set of optimal pseudo-gaps $\gamma_k$. For sufficiently large systems, the exponential growth of the number of spectral gaps practically ensures that the spectrum $\rho(\Delta)$ can be viewed as a continuous distribution of $\Delta$ presented in Fig. \ref{fig:gap-selection}. Such distribution is limited to the range $[0,\Delta_{\rm{max}}]$ and have a decreasing tail in the region of large $\Delta$ values due to combinatorial calculation of spectral gaps from the generator's eigenvalues.. When using GPSR, all spectral gaps from this distribution contribute to the derivative computation. This can be interpreted as sampling the gap distribution $\rho(\Delta)$ with a very small step $\rm{d}\Delta\rightarrow0$. Recall that in the case of aGPSR we have demonstrated in subsection \ref{subsec: deriv} that one can use an arbitrary set of pseudo-gaps $\gamma_k$, and still get an increasingly accurate estimate of the derivative. Therefore, to minimize the error function $Q_K(\Delta)$ the pseudo-gaps should represent the real ones as closely as possible, or in other words $\gamma_k$ should be sampled from the real-gap distribution $\rho(\Delta)$, preferably from high-probabilty regions. Therefore, the selected $K$ values of $\gamma_k$ would be optimal for the most accurate estimate of the derivative. However, such sampling is too costly to implement for larger systems, as the eigenspectrum scales exponentially. 

On the other hand, we could still use efficient algorithms \cite{lehoucq1998arpack} to estimate the largest and smallest eigenvalues of the generator $\hat{G}$ to calculate the largest real spectral gap $\Delta_{\rm{max}}$. This value at least gives information about the extent of the spectral gap distribution. Without \textit{a priori} knowledge of $\rho(\Delta)$ one could resort to sampling selecting $K$ pseudo-gaps $\gamma_k$ uniformly and equidistantly in the interval $[0, \Delta_{\rm{max}}]$ with some step $a$. Intuitively, this method allows us to cover the unknown distribution of spectral gaps $\Delta$ consistently and independently of the actual shape of it. In practice it means that one can set beforehand the value for $K$, e. g., according to computational requirements on algorithm execution speed, and then vary the step $a$ to achieve the best estimate of the derivative. Increasing $a$ covers more area of the spectral gap distribution $\rho(\Delta)$, however smaller $a$ allows to focus on the initial part of the distribution where often the most real spectral gaps are found. Such procedure is illustrated in Fig. \ref{fig:gap-selection}.

\section{Variance estimation}
\label{variance_est}

The application of aGPSR on real quantum devices introduces an additional aspect affecting the accuracy of estimated derivatives - the probabilistic nature of quantum mechanical measurement process that is used to calculate the expectation value of some cost operator, hence the function $f(x)$. For a finite number of measurements $N_{\rm shots}$ each measurement results in a random value of $f(x)$ that we assume follows normal probability distribution characterized by mean value $\bar{f}(x)$ and variance $\frac{\sigma^2_0(x)}{N_{\rm shots}}$. Correspondingly, the derivative $\frac{{\rm d}\bar{f}\left(x\right)}{{\rm d}x}$ also depends on the number of included equations in the linear system \ref{Rs_eq_system} and chosen shift values $\delta$ through shifted functions $f(x\pm\delta)$. It is therefore relevant to investigate the minimization of the variance of the derivative estimation by optimally selecting shift values.

To answer this, let us first consider the full linear equation system in Eq.~\ref{Rs_eq_system} representing exact GPSR and rewrite the definition of the derivative $\frac{{\rm d}f\left(x\right)}{{\rm d}x}$ to emphasize the dependence on shifted function values:
\begin{equation}
\frac{{\rm d}f\left(x\right)}{{\rm d}x}=\overset{S}{\underset{s=1}{\sum}}\Delta_{s}R_{s}(F_1, F_2, ..., F_S). \label{dfdx_definition_expl_dep}
\end{equation}
Since we want to calculate the variance $\sigma^2(x)$ of the derivative above we must find explicit expressions for $R_{s}(F_1, F_2, ..., F_S)$. In subsection \ref{subsec: deriv} we solved the linear system \ref{Rs_eq_system} with Cramer's rule, however for current derivation it will be easier to consider matrix-vector equation \ref{matrix_eq_system} and write the solution for $\mathbf{R}^{S\times 1}$ as 
\begin{equation}
\mathbf{R}^{S\times 1} = \left(\mathbb{M}^{S\times S}\right)^{-1}\mathbf{F}^{S\times 1}.\label{matrix_eq_system_solution}
\end{equation}
From this expression, we can see that $R_{s}(F_1, F_2, ..., F_s)$ linearly depends on all the differences $F_s$, thus it can be written in the following way:
\begin{equation}
R_{s}(F_1, F_2, ..., F_S) = \overset{S}{\underset{k=1}{\sum}}\left(\mathbb{M}^{S\times S}\right)^{-1}_{sk}F_k = \overset{S}{\underset{k=1}{\sum}}a_{sk}(f(x+\delta_k)-f(x-\delta_k)).\label{Rs_inv_mat_sol}
\end{equation}
As we pointed out in previous paragraphs, $f(x\pm\delta_k)$ is a stochastic quantity when evaluated with some measurement protocol on a real quantum device, thus $R_{s}(F_1, F_2, ..., F_S)$ in Eq. \ref{Rs_inv_mat_sol} can be interpreted as a sum of scaled independent random values. Following \cite{gpsr2021} we also assume that the stochastic values $f(x\pm\delta_s)$ are identically distributed and the dependence on the argument value can be neglectes. Thus, their variances are constant - $\sigma^2_0(x)\approx\sigma^2_0$. This way, we can write the variance of $R_s$ that is a sum of iid stochastic variables in Eq. \ref{Rs_inv_mat_sol} as a sum of scaled variances of individual stochastic variables:
\begin{equation}
\sigma^2_s = \frac{2\sigma^2_0}{N_{\rm shots}}\overset{S}{\underset{k=1}{\sum}}a^2_{sk}.\label{Rs_variance}
\end{equation}
However, from the derivative given by Eq. \ref{dfdx_definition_expl_dep} and the expression for variance of $R_s$ in Eq. \ref{Rs_variance}, we can finally write the variance of the derivative $\sigma^2_{\rm d}$ in the following way:
\begin{equation}
\sigma^2_{\rm d} = \overset{S}{\underset{s=1}{\sum}}\Delta^2_s \sigma^2_s = \frac{2\sigma^2_0}{N_{\rm shots}}\overset{S}{\underset{s=1}{\sum}}\overset{S}{\underset{k=1}{\sum}}\Delta^2_s a^2_{sk}.\label{dfdx_variance}
\end{equation}
Since coefficients $a_{sk}$ are actually components of the inverted matrix $\left(\mathbb{M}^{S\times S}\right)^{-1}$, they are functions of all the spectral gaps $\{\Delta_s\}$ and shifts $\{\delta_s\}$. This means that we can minimize the function
\begin{equation}
g(\{\Delta_s\}, \{\delta_s\})=\overset{S}{\underset{s=1}{\sum}}\overset{S}{\underset{k=1}{\sum}}\Delta^2_s a^2_{sk}(\{\Delta_s\}, \{\delta_s\}) \label{var_minimize}
\end{equation}
with respect to the shifts $\{\delta_s\}$ for a given set of spectral gaps $\{\Delta_s\}$ and obtain the optimal set of parameter shift values $\{\delta^{\rm opt}_s\}$ that results in the smallest variance of the derivative given by Eq. \ref{dfdx_definition_expl_dep} when estimated on a real quantum device. This optimization procedure can also be applied with aGPSR when we consider only a part of equations in system \ref{Rs_eq_system}. In this case, the spectral gaps are chosen as described in subsection \ref{subsec: gap-select} to minimize the error resulting from truncated equation system and then with the optimal spectral gaps fixed the shift minimization can be performed as usual. 

Finally, the procedure for selecting optimal, variance-minimizing parameter shift values for aGPSR can be summarized in the following way:
\begin{enumerate}
\item Perform spectral gap optimization to find the set of gaps that minimize the derivative error for a given number of included equations $K$.
\item Calculate the analytical inverse of matrix $\mathbb{M}^{K\times K}$ to obtain coefficients $a_{sk}(\{\Delta_s\}, \{\delta_s\})$.
\item Minimize $g(\{\Delta_s\}, \{\delta_s\})=\overset{K}{\underset{s=1}{\sum}}\overset{K}{\underset{k=1}{\sum}}\Delta^2_s a^2_{sk}(\{\Delta_s\}, \{\delta_s\})$ with respect to shifts $\{\delta_s\}$.
\item Use the optimal spectral gaps together with optimal shift values to estimate derivative on a real quantum device with minimal bias and variance.
\end{enumerate}

\section{Results}
\label{results}

In the next sections, we demonstrate the suitability of aGPSR for derivative calculations and its efficiency when applied within a variational quantum eigensolver (VQE) task.

\subsection{Derivative calculations}

\begin{figure}[H]
    \centering
    \begin{subfigure}{0.45\textwidth}
        \centering
        \includegraphics[width=\textwidth]{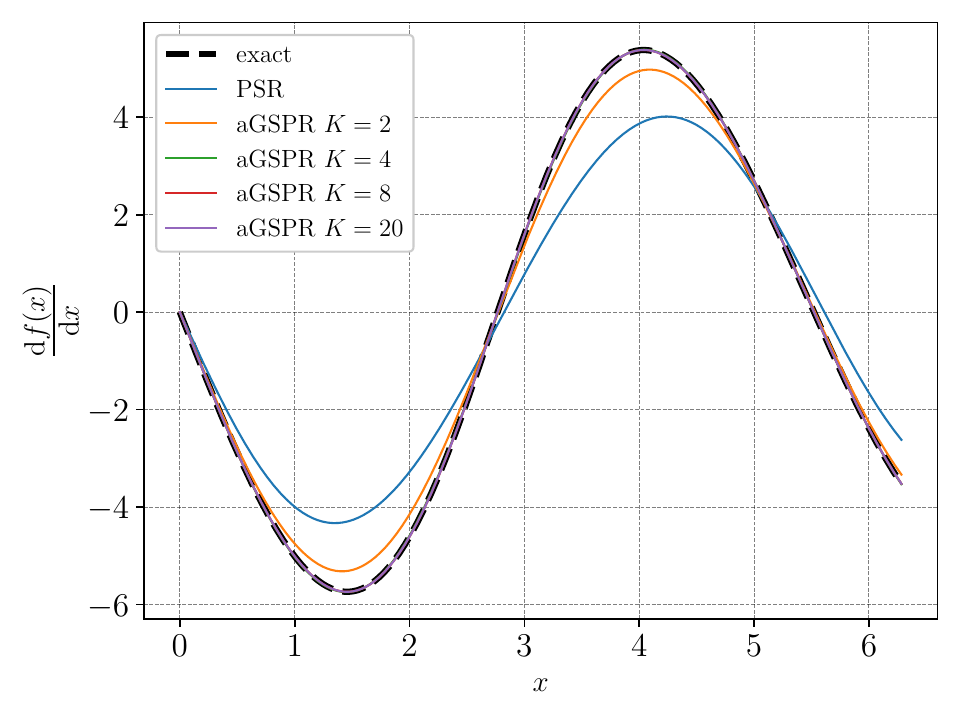}
        \caption{Ratio of the drive amplitude with the interaction strength is $\frac{J}{\Omega} < 1$.}
        \label{fig:weak-int-zero}
    \end{subfigure}
    \hfill
    \begin{subfigure}{0.45\textwidth}
        \centering
        \includegraphics[width=\textwidth]{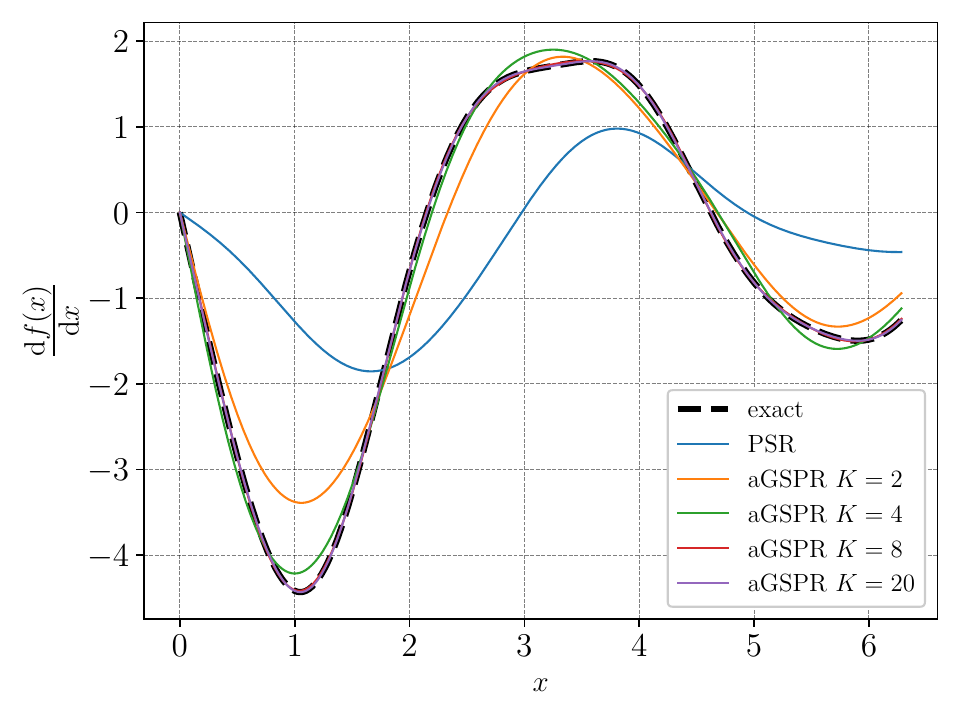}
        \caption{Ratio of the drive amplitude with the interaction strength is $\frac{J}{\Omega} > 1$.}
        \label{fig:strong-int-zero}
    \end{subfigure}
    \caption{Comparison of derivatives $\frac{{\rm d}f}{{\rm d}x}$ calculated for different $K$ values with zero initial state $\left|\psi_0\right\rangle$.}
    \label{fig:zero-init-state}
\end{figure}

The accuracy of aGPSR and its suitability for calculating derivatives on quantum systems exhibiting arbitrary qubit interactions can be illustrated by considering the neutral-atom Hamiltonian \cite{henriet2020}:
\begin{equation}
\hat{H}=\sum_{i=1}^N\frac{\Omega}{2}\hat{\sigma}_{i}^{x}+\sum_{j<i}^N J_{ij}\hat{n}_{i}\hat{n}_{j}. \label{neutr-atom-ham}
\end{equation}
Here, $\Omega$ is the drive amplitude of the laser pulse, $J_{ij}$ is the interaction strength between atoms $i$ and $j$ and $\hat{n_i}=\frac{\hat{\sigma}_{i}^{z} + \hat{I}_i}{2}$ the number operator. The unitary evolution operator in this case can be written as 
\begin{equation}
\hat{U(t)}={\rm exp}\left(-it\hat{H}\right)={\rm exp}\left(\frac{-ix}{2}\hat{G}\right), \label{neutr-atom-gen}
\end{equation}
where the differentiation parameter is $x=\Omega t$ and the corresponding neutral-atom generator is $\hat{G}$. In Fig. \ref{fig:zero-init-state} we can see the derivative calculated for a 6-qubit system with a $2\times3$ square qubit-layout and a zero initial state $\left|\psi_0\right\rangle=\left|0\right\rangle^{\otimes 6}$. Parameters of the generator $\hat{G}$ were selected such that either $\Omega$ is approximately 2 times smaller than the characteristic interaction strength $J$ or larger. In the former case, the interaction between qubits is weak and the dynamics of the system are mainly determined by the drive amplitude. However, as we can see from Fig. \ref{fig:weak-int-zero}, even this weak interaction is enough for ordinary PSR (blue curve) to give quite inaccurate derivative estimate because the generator of the system cannot be represented as tensor product of single-qubit operations. Using aGPSR under the same conditions allows for a perfect reconstruction of the derivative when using $K=4$ equations with pseudo-gap step $a=4$. In contrast, the full GPSR algorithm for a 6-qubit system would require to solve a linear system with $S=2016$ spectral gaps and equations. Even in this reduced example, savings in computational cost are massive. 

Looking at the case in Fig. \ref{fig:strong-int-zero} the inadequacy of PSR is even more evident since the interaction between the qubits now dominates and determines the behavior of the system. However, even under these more challenging conditions aGPSR with $K=8$ already provides a perfect match with the exact derivative function. Note, that with $K=4$, the discrepancy between exact result and the aGPSR is quite significant, meaning that larger interaction leads to a more extended spectral gap distribution that requires more points to cover properly and obtain an adequate derivative estimate. 

\begin{figure}[H]
    \centering
    \begin{subfigure}{0.45\textwidth}
        \centering
        \includegraphics[width=\textwidth]{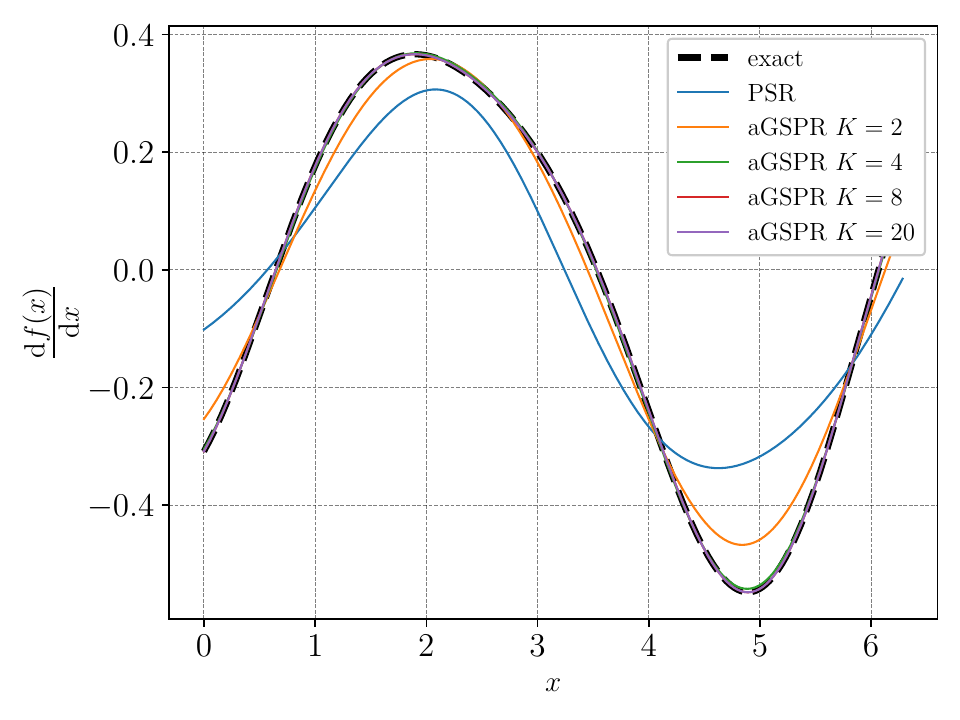}
        \caption{Ratio of the drive amplitude with the interaction strength is $\frac{J}{\Omega} < 1$.}
        \label{fig:weak-int-rand}
    \end{subfigure}
    \hfill
    \begin{subfigure}{0.45\textwidth}
        \centering
        \includegraphics[width=\textwidth]{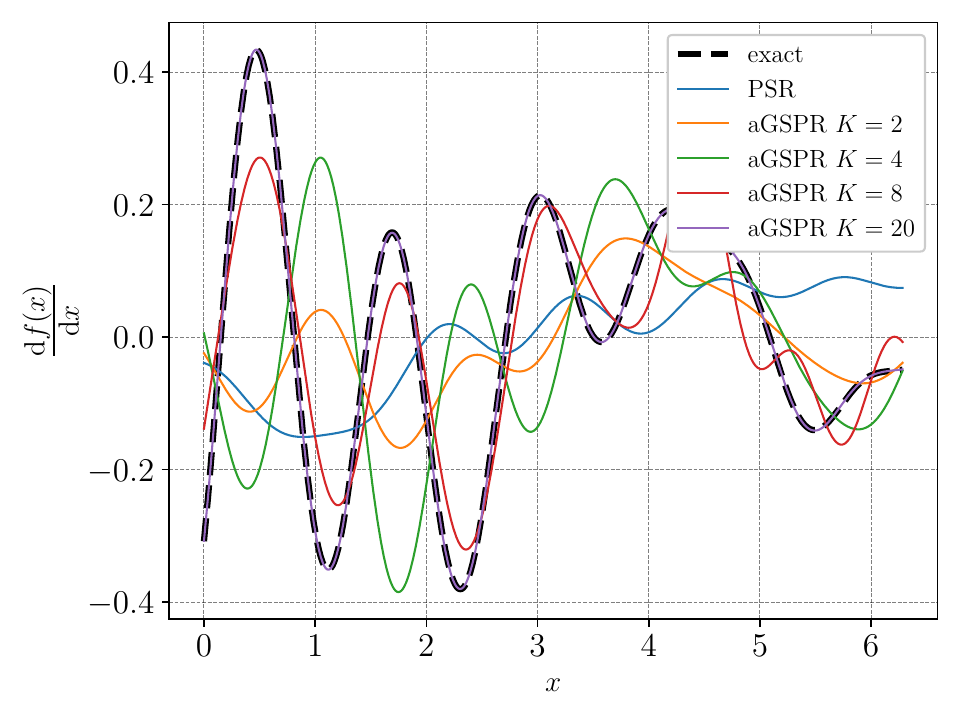}
        \caption{Ratio of the drive amplitude with the interaction strength is $\frac{J}{\Omega} > 1$.}
        \label{fig:strong-int-rand}
    \end{subfigure}
    \caption{Comparison of derivatives $\frac{{\rm d}f}{{\rm d}x}$ calculated for different $K$ values with random initial state $\left|\psi_0\right\rangle$.}
    \label{fig:rand-init-state}
\end{figure}

From the results of aGPSR calculations for a random initial state presented in Fig.~\ref{fig:rand-init-state}, we can see that in the same manner as with zero initial state, the weak interaction setup requires $K=4$ equations for aGPSR to deliver perfect match with the exact derivative. On the other hand, within the strong interaction regime, a significantly larger number $K=20$ is necessary to achieve the best match. Intuitively, the reason stems from the fact that initial state randomness rules out any distinctive features in the spectral gap distribution, \textit{e. g.}, a high probability region where the bulk of spectral gaps distributed, thus requiring a finer sampling coverage.

\begin{figure}[H]
    \centering
    \begin{subfigure}{0.45\textwidth}
        \centering
        \includegraphics[width=\textwidth]{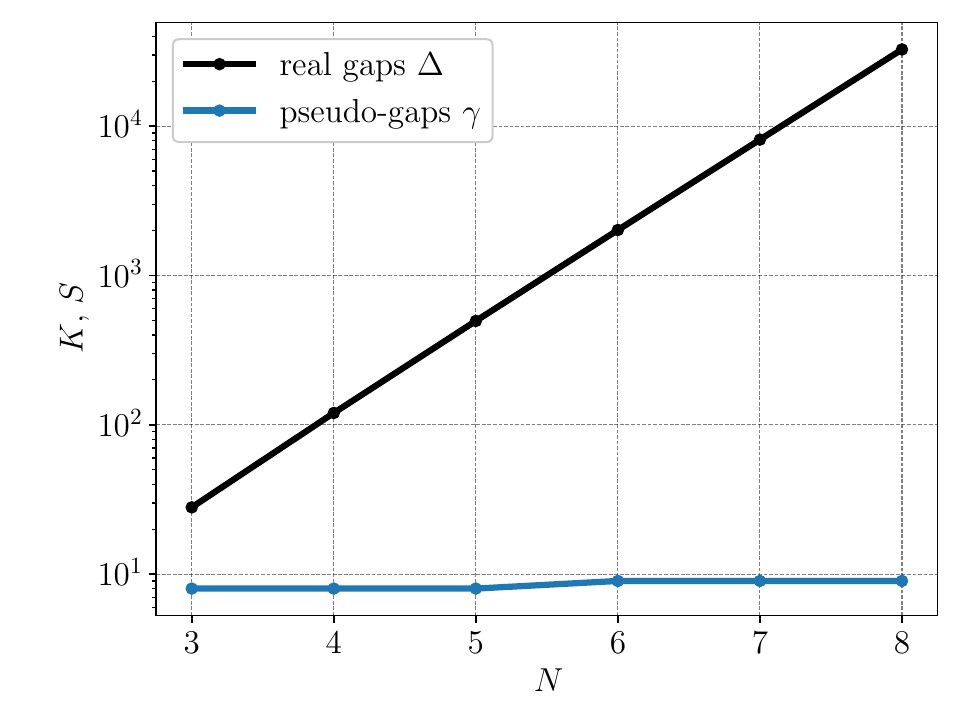}
        \caption{Scaling of the required number real gaps $\Delta$ and pseudo-gaps $\gamma$ to achieve relative error $r\le0.2\%$ with system size $N$.}
        \label{fig:num-gaps-scaling}
    \end{subfigure}
    \hfill
    \begin{subfigure}{0.45\textwidth}
        \centering
        \includegraphics[width=\textwidth]{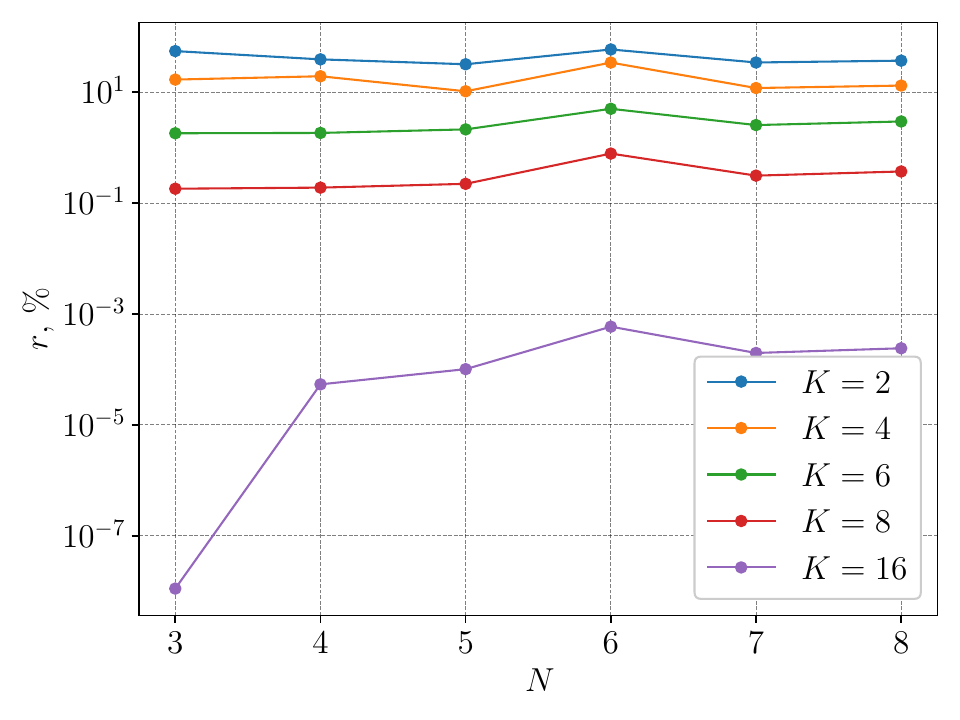}
        \caption{Scaling of the relative error $r$ with system size $N$ for different numbers $K$ of equations used with aGPSR.}
        \label{fig:re-scaling}
    \end{subfigure}
    \caption{Comparison of derivatives $\frac{{\rm d}f}{{\rm d}x}$ calculated for different $K$ values with zero initial state $\left|\psi_0\right\rangle$.}
    \label{fig:scaling}
\end{figure}

Let us investigate the scaling of the number $K$ denoting the minimum required number of equations in aGPSR framework to achieve some target mean relative error $r$ of the derivative estimate. We define the relative error as 
\begin{equation}
r=\frac{1}{D}\sum_{i=1}^{D}\frac{|f^{\prime}_{\rm{aGPSR}}(x_i)-f^{\prime}_{\rm{exact}}(x_i)|}{f^{\prime}_{\rm{exact}}(x_i)}, \label{eq:rel-error-comp}
\end{equation}
where $f^{\prime}_{\rm{aGPSR}}(x_i)$ denotes the aGPSR estimate of the derivative at point $x_i$ and $f^{\prime}_{\rm{exact}}(x_i)$ is the exact value of the derivative. The value $D$ denotes the number of derivative samples taken to calculate the mean value. In Fig. \ref{fig:num-gaps-scaling} we demonstrate the scaling of the required computational effort represented by the required number of gaps for both full GPSR and aGPSR for a system where the interaction between qubits $J$ is approximatelly equal to the drive amplitude $\Omega$. We can see that the total number of gaps $\Delta$ increases exponentially with system size, however with aGPSR to achieve the value of relative error $r\approx0.2\%$ the number of equations $K$ is practically constant for each number of $N$. This striking difference in scaling of the number of gaps between GPSR and aGPSR is a good confirmation of the practicality of the latter. In Fig. \ref{fig:re-scaling} we can see the dependencies of the relative error $r$ on the selected number of equations $K$ in aGPSR and the system size $N$. These results confirm the fact that larger $K$ values lead to smaller derivative estimation errors. We can also notice that with the exception of $K=16$, the relative error is essentially constant for every value of the system size $N$. 

\subsection{Variational quantum eigensolver}

In this section, we demonstrate the efficiency of aGPSR when applied within a variational quantum eigensolver (VQE) task \cite{Peruzzo2014}. A VQE task consists in preparing a quantum state or ansatz defined by a set of parameters $\theta$ denoted $\ket{\psi(\theta)}$, minimizing the energy with respect to a given Hamiltonian $H$ : $\bra{\psi(\theta)} H \ket{\psi(\theta)}$. The parameters $\theta$ are tweaked by a classical optimizer, commonly a gradient-based one such as Adam.  

\begin{figure*}[!ht]
\centering
\subfloat[]{\includegraphics[width=0.8\columnwidth]{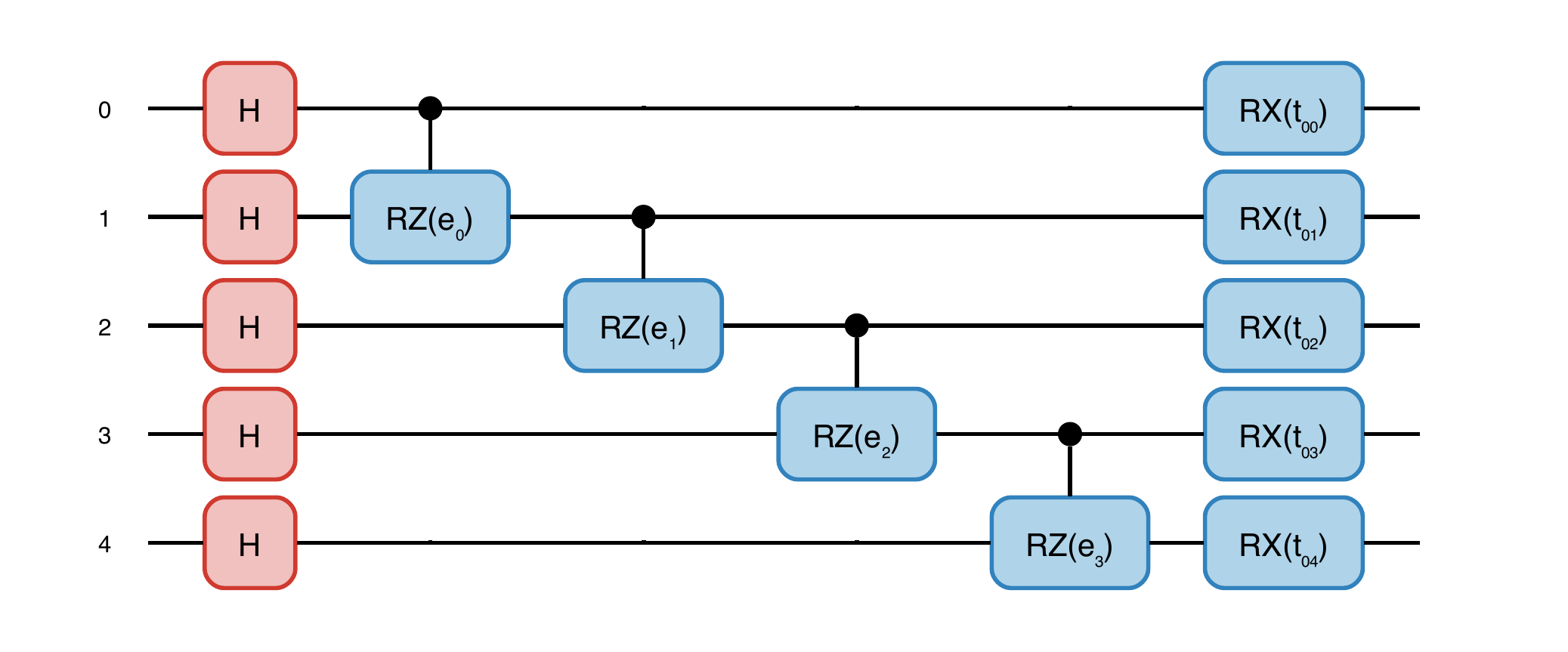}}
\\
\subfloat[]{\includegraphics[width=0.45\columnwidth]{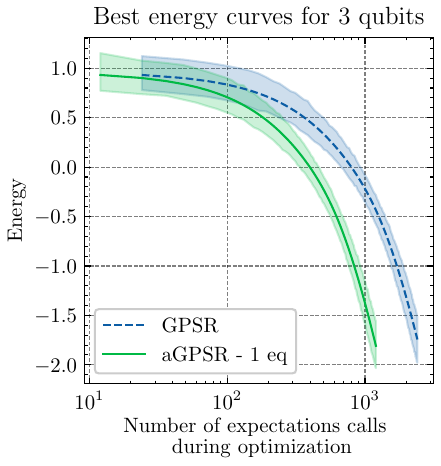}}
\subfloat[]{\includegraphics[width=0.43\columnwidth]{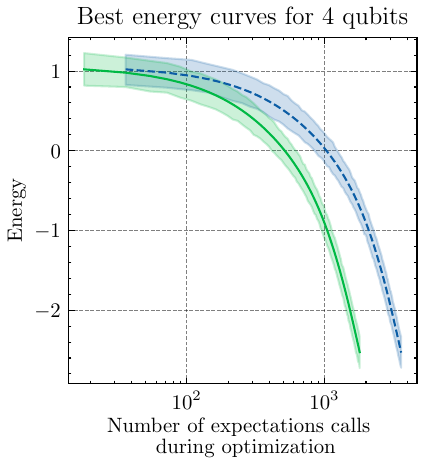}}
\\
\subfloat[]{\includegraphics[width=0.43\columnwidth]{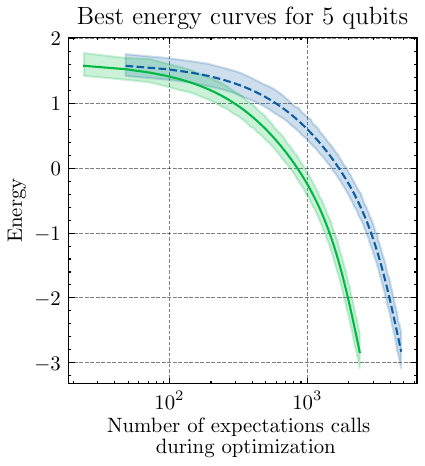}}
\subfloat[]{\includegraphics[width=0.43\columnwidth]{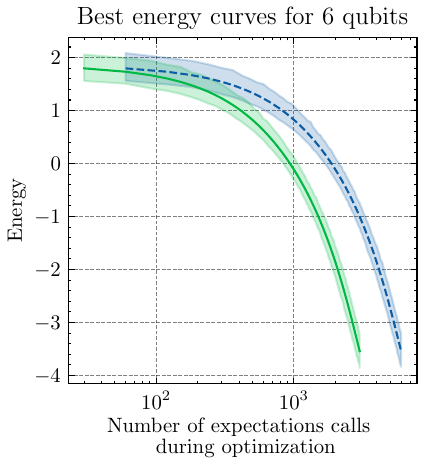}}
\caption{VQE experiments ($10$ runs per differentiation method) using a digital ansatz (we use a circuit consisting of $3$ repetitions of the layer shown in (a).) for $3$ (b), $4$ (c), $5$ (d), $6$ (e) qubits.}
\label{vqseperfsdigital}
\end{figure*}

We study two test cases of ansatzes: the first one is defined by a digital quantum circuit (composed of $3$ repetitions of the layer shown in Fig.~\ref{vqseperfsdigital} (a)) while the second is composed of a single analog operation (the same strong interaction case described in Section~\ref{results}). In our experiments, the Hamiltonian of interest is defined as $\sum_{i=1}^N Z_i$ where $N$ is the number of qubits, varying from $3$ to $6$. Our experiments do not involve shots, but rather we optimize the analytical expectation value. The underlying optimizer, Adam, is run for $100$ iterations with a learning rate of $0.01$. We perform $10$ VQE runs per differentiation method and system. 

\begin{figure*}[!ht]
\centering
\subfloat[]{\includegraphics[width=0.45\columnwidth]{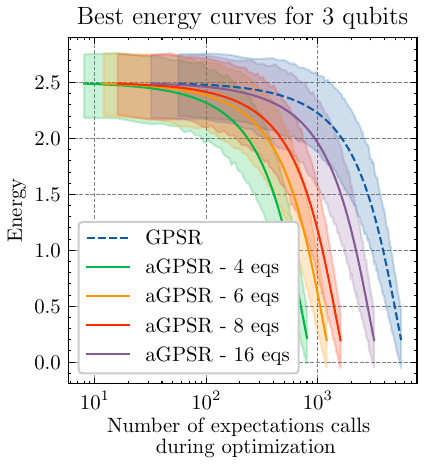}}
\subfloat[]{\includegraphics[width=0.43\columnwidth]{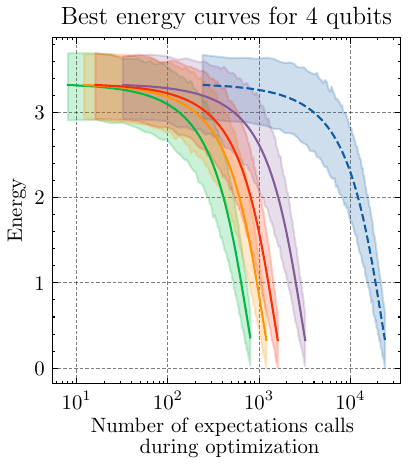}}
\\
\subfloat[]{\includegraphics[width=0.43\columnwidth]{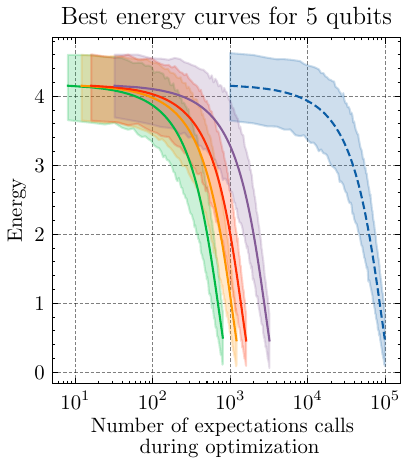}}
\subfloat[]{\includegraphics[width=0.43\columnwidth]{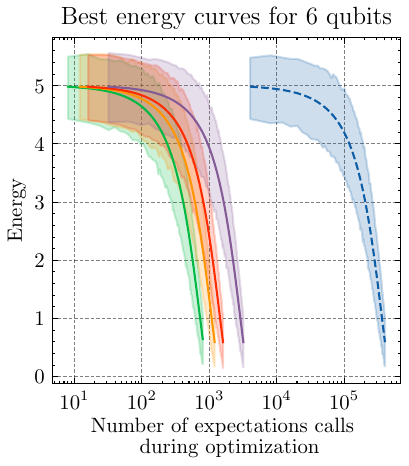}}
\caption{VQE experiments ($10$ runs per differentiation method) using an analog ansatz (using a single analog operation described in Section~\ref{results}) for $3$ (a), $4$ (b), $5$ (c), $6$ (d) qubits.}
\label{vqseperfsanalog}
\end{figure*}

As shown in Fig.~\ref{vqseperfsdigital}, we decrease the runtime of VQE, depicted as the number of expectation calls $S$ involved when running the multi-gap GPSR (as defined in Eq.~\ref{Rs_eq_system}), while matching the best energy found using the Adam optimizer by using $K=1$ equations with aGPSR instead of $2$ using GPSR. However, in the analog ansatz case (Fig.~\ref{vqseperfsanalog}), the saved factor is significantly more pronounced as the number of equations involved with GPSR increases ($S=\frac{2^N (2^N-1)}{2}$, corresponding respectively to $28$, $120$, $496$, and $2016$ from $3$ to $6$ qubits). Indeed, the factor of expectation calls saved when $K=4$ is respectively $7, 30, 124$ and $504$. Thus, aGPSR is relevant for saving runtime when performing expensive variational quantum algorithms training.

These results were produced using Pasqal's open-source quantum software development kit Qadence \cite{qadence2025} where an aGPSR implementation can be found.

\section{Discussion and conclusions}


Whilst GPSR is mathematically exact, it practically fails to evaluate an exponential growing number of shifted points for limited shot-budget experiments. aGPSR addresses this shortfall by retaining GPSR's ability to handle arbitrary generators together with providing flexibility to adjust the accuracy of the estimated derivative. Since with aGPSR the linear equation system required to be solved is much smaller than with GPSR, one needs much smaller number of function $f(x)$ evaluations on a real device or emulator to estimate derivative, thus drastically reducing the necessary compute power to achieve an accurate result in the end. Small numbers of pseudo-gaps used in aGPSR calculations also facilitates the search for optimal set of parameter shifts that minimize the variance of the derivative estimation when applying the method on a realistic device where the shot budget of function evaluations is limited. We demonstrated its usefulness on a VQE task defined over $3$ to $6$ qubits where the number of functions evaluations saved range from $7$ to $504$, for the same target reached.

As future work, we can explore applying aGPSR for different applications, in noisy settings, and on real device. Different strategies can be designed for choosing the pseudo-gaps, involving techniques from classical optimization to machine learning. Indeed, in the context of variational quantum algorithms, we can make the gap selection adaptive when performing gradient descent where the first iterations would use a tiny number of gaps, while later ones increase the number. We can also study the usage of classical surrogates to limit the calls to quantum devices \cite{surrogate}. Finally, we can study possible combinations of aGPSR with shot-frugal optimizers, especially relevant in the context of Quantum Machine Learning applications\cite{refoqus}.

\subsection*{Ethics declaration}
Pasqal has filed a patent for aGPSR~\cite{agpsr-patent}.

\bibliographystyle{unsrt}  
\bibliography{references}

\end{document}